\documentclass[useAMS,usenatbib]{mn2e}

\usepackage{enumerate}
\usepackage{graphicx}
\usepackage{xcolor} 

\newcommand{\simgt}{\lower.5ex\hbox{$\;\buildrel>\over\sim\;$}}
\newcommand{\simlt}{\lower.5ex\hbox{$\;\buildrel<\over\sim\;$}}

\newcommand{\msun}{\ensuremath{M_\odot}}
\newcommand{\msunyr}{$M_\odot$\,yr$^{-1}$}
\newcommand{\ergs}{erg\,s$^{-1}$}

\newcommand{\mstar}{\ensuremath{M_{\rm star}}}

\newcommand{\hii}{\rm H{\sc ii}}

\newcommand{\ha}{H$\alpha$}
\newcommand{\hb}{H$\beta$}
\newcommand{\bra}{Br$\alpha$}
\newcommand{\te}{T$_{\rm e}$}

\newcommand{\logoh}{12$+\log$(O/H)}

\newcommand{\lir}{$L_{\rm IR}$}

\newcommand{\sigmasfr}{$\Sigma_{\rm SFR}$}
\newcommand{\sbs}{SBS\,0335$-$052}
\newcommand{\izw}{I\,Zw\,18}
\newcommand{\spit}{{\it Spitzer}}
\newcommand{\oiii}{\rm [O{\sc iii}]}
\newcommand{\oii}{\rm [O{\sc ii}]}

\newcommand{\aap}{A\&A}
\newcommand{\apj} {ApJ}
\newcommand{\apjl} {ApJL}
\newcommand{\apjs} {ApJS}
\newcommand{\aj} {AJ}
\newcommand{\araa} {ARA\&A}
\newcommand{\mnras}{MNRAS}

\newcommand{\pasp}{PASP}

\title[Scaling relations of metal-poor starbursts]{Scaling relations of
metallicity, stellar mass, and star formation rate in metal-poor
starbursts: I. A fundamental plane}

\author[L.~K. Hunt et al.]{Leslie Hunt$^{1}$\thanks{E-mail: hunt@arcetri.astro.it},
Laura Magrini$^{1}$, Daniele Galli$^{1}$, Raffaella Schneider$^{2}$, Simone Bianchi$^{1}$,
\newauthor 
Roberto Maiolino$^{3}$,
Donatella Romano$^{4}$,
Monica Tosi$^{4}$,
Rosa Valiante$^{2}$\\
$^{1}$INAF/Osservatorio Astrofisico di Arcetri, Largo Enrico Fermi 5, 50125 Firenze, Italy\\
$^{2}$INAF/Osservatorio Astronomico di Roma, Via di Frascati 33, 00040 Monteporzio, Italy\\
$^{3}$Cavendish Laboratory, University of Cambridge, 19 JJ Thomson Avenue, Cambridge CB3 0HE, UK\\
$^{4}$INAF/Osservatorio Astronomico di Bologna, Via Ranzani 1, 40127 Bologna, Italy}
\begin{document}

\date{}

\pagerange{\pageref{firstpage}--\pageref{lastpage}} \pubyear{2012}

\maketitle

\label{firstpage}

\begin{abstract}
Most galaxies follow well-defined
scaling relations of metallicity (O/H), star formation rate (SFR), and
stellar mass (\mstar).  However, low-metallicity starbursts, 
rare in the Local Universe but more common at high redshift, deviate 
significantly from these scaling relations.
On the ``main sequence'' of star formation, these galaxies have high SFR for a given 
\mstar; and on the mass-metallicity relation, they have excess \mstar\ for 
their low metallicity.  
In this paper, we characterize O/H,
\mstar, and SFR for these deviant ``low-metallicity starbursts'',
selected from a sample of $\sim$1100 galaxies, spanning almost two orders of
magnitude in metal abundance, a factor of $\sim10^6$ in SFR, and of
$\sim10^5$ in stellar mass.
Our sample includes quiescent star-forming galaxies and 
blue compact dwarfs at redshift 0, luminous compact galaxies at
redshift 0.3, and Lyman Break galaxies at redshifts 1-3.4. 
Applying a Principal Component Analysis (PCA) to the
galaxies in our sample with \mstar$\leq3\times10^{10}$\,\msun\ 
gives a Fundamental Plane (FP) of scaling relations; 
SFR and stellar mass define the plane itself, and O/H its thickness.
The dispersion for our sample in the edge-on view of the plane is 0.17\,dex,
independently of redshift and including the metal-poor starbursts.
The same FP is followed by 55\,100 galaxies selected
from the Sloan Digital Sky Survey, with a dispersion of 0.06\,dex.
In a companion paper, we develop multi-phase chemical evolution models that 
successfully predict the observed scaling relations and the FP;
the deviations from the main scaling relations are caused by a different 
(starburst or ``active'') mode of star formation.
These scaling relations do not truly evolve, but rather are defined by
the different galaxy populations dominant at different cosmological epochs.
\end{abstract}


\section{Introduction} \label{sec:intro}

Our understanding of how galaxies assemble their mass over cosmic time has 
advanced considerably in the last decade.
There have emerged observationally-defined scaling relations that link
metallicity, stellar mass, and star formation rates (SFRs) in galaxies 
for lookback times of up to $\sim$10\,Gyr. 
The rate at which galaxies
form stars is correlated with their stellar mass,
up to $\ga 10^{11}$\msun where the trend flattens.
This trend, known as the ``main sequence of star formation'' (SFMS) has roughly
the same slope up to redshifts $z\sim$3, but an increasing normalization
such that higher redshift galaxies produce more stars for their mass
than low-redshift ones
\citep{noeske07,salim07,schiminovich07,bauer11}.
Thus, high-redshift populations tend to have higher
specific SFRs (SFR divided by stellar mass, sSFR) than nearby galaxies, 
although there is still
some debate about whether at all redshifts
higher sSFRs are observed in less massive galaxies than in their 
more massive counterparts 
\citep[e.g.,][]{zheng07,perez08,rodighiero10}. 

Metallicity and stellar mass are also related (the mass-metallicity relation,
MZR), but through 
a somewhat tighter correlation 
\citep{tremonti04,savaglio05,erb06,maiolino08,mannucci09}. 
A secondary dependence of the MZR on SFR, dubbed the ``Fundamental
Metallicity Relation'' (FMR), reduces the scatter considerably for local
galaxy populations \citep{mannucci10}. 
\citet{mannucci11} extended the FMR
to lower stellar masses with Gamma-Ray-Burst host galaxies, but it 
is still unable to encompass the properties
of galaxies at $z\ga 3$, and as we show below, also in nearby metal-poor
starbursts.

Although the SFMS and the MZR (and FMR) are good at predicting the
links among star formation, metallicity, and stellar mass for
the bulk of galaxies up to redshifts of $\sim$2, some subsets of
galaxy populations show clear departures from the main trends.
In particular, vigorous starbursts with high SFRs tend to lie
above the SFMS at all redshifts and to have too much mass (or luminosity)
for their metallicity \citep[e.g.,][]{hoyos05,rosario08,salzer09,peeples09,rodighiero11,atek11}.
Usually, these high-SFR outliers are attributed to the merger mode
of star formation; galaxy-wide SF occurring over long dynamical times   
is thought to define the SF Main Sequence, while 
{\it merger-induced bursts} of star formation result in outliers from the main trend
\citep[e.g.,][]{rodighiero11,lee04}.

In a companion paper \citep{magrini12}, we propose 
that mergers may not be directly responsible for producing
the starbursts that deviate from the general SFR-mass-metallicity relations,
but rather that initial physical conditions, independently of their origin, may
be driving the deviations.
We develop different classes of models based on distinct sets of initial conditions,
including mass and initial density of the galaxy, and size and density of the star-forming regions.
These would correspond to different modes of star formation,
an ``active'' or starburst mode, and a more ``passive'', quiescent one, with
which we successfully 
model the scaling relations of $\la$1100 galaxies.
Here, we present the sample of galaxies, and describe the observed scaling relations
of metallicity, stellar mass, and star formation rate for the sample.

\subsection{Our approach} 

In this paper, we want
to investigate the empirical connection between high-$z$ starbursts and
metal-poor galaxies with high SFRs in the Local Universe.
Our aim is not to analyze the bulk of the star-forming galaxies representative
of the Local Volume, as has already been done by several groups with the SDSS
\citep[e.g.,][]{tremonti04,mannucci10,laralopez10,yates12,zahid12}.
Rather, our intention is to identify classes of galaxies which deviate from
the SFMS and the MZR in the same way, placing particular emphasis on 
potential local analogues of high-redshift galaxy populations. 
The idea is that the outliers of well-defined scaling relations may reveal
insights that cannot be gained otherwise.
Hence, we require a sample that spans a wide range in parameter space, in order
to adequately characterize trends with metallicity, SFR, and stellar mass
over a large redshift interval.

As we show below,
some low-metallicity Blue Compact Dwarf galaxies (BCDs) have high SFRs, 
particularly high sSFRs, and occupy the same locus of the MZR and the SFMS
as $z=3$ Lyman Break Galaxies (LBGs). Relative to the main trend,
these objects have an excess of $\ga$100 in stellar mass at a given metallicity,
and a factor of $\sim10$ excess SFR at a given mass.
The same is true for some of the ``Green Peas'' and Luminous Compact Galaxies (LCGs) at $z\sim0.1-0.4$
selected by \citet{cardamone09} and \citet{izotov11} from the Sloan Digital Sky Survey (SDSS).
It seems evident that SFR (or SFR surface density \sigmasfr, which
is more difficult to measure) is governed by similar physical conditions
at all redshifts, and that which galaxy populations are observed at
a given redshift depends on how they are selected. 
Because extreme starbursts are rare locally, but more common at high redshift, 
selection effects\footnote{We use the term ``selection effect''
to mean ``the tendency for a conclusion based on observations to be influenced by
the method used to select objects for observation'' 
(see {\it http://earthguide.ucsd.edu/virtualmuseum/Glossary\_Astro/\\
gloss\_s-z.shtml}).
In this sense, how a galaxy (or galaxy population) is selected may be important for
defining the slope and normalization of scaling relations, both locally and at high redshift.} 
may be paramount in defining observed trends among
stellar mass, metallicity, and SFR.

The paper is organized as follows:
in Sect.~\ref{sec:samples} we present several samples of dwarf galaxies in
the Local Universe, two samples of intermediate-redshift galaxies
(the ''Green Peas'' and LCGs), and several samples of LBGs at $z\sim1-3$.
We also explain how we derive stellar masses from 4.5\,\micron\
observations for the nearby galaxy samples. 
Sect.~\ref{sec:lms} identifies a class of ``low-metallicity starbursts''
on the basis of deviations from the main scaling relations observed at
low and high redshift.
A ``Fundamental Plane'' of metallicity, stellar mass, and star formation rate
independent of redshift
is introduced in Sect.~\ref{sec:fp}, and its characteristics are compared
with common perceptions of the SFMS and the MZR.
In Sect.~\ref{sec:conclusions}, 
we discuss possible similarities between nearby starbursts 
and starbursts at high redshift,
and possible ramifications for our current conception 
of galaxy evolution. 

\medskip
\section{Observational samples} \label{sec:samples}

Because of the need to compare SFR, metal abundance [as defined by the
nebular oxygen abundance, \logoh], and stellar mass, \mstar,
we have selected only samples of galaxies which either have these quantities
already available in the literature, or for which we could derive them
from published data.
Five sets of 
galaxies in the Local Universe met these criteria:
the nearby dwarf irregular galaxies (dIrr) published by \citet[][here dubbed ``dIrr'']{lee06},
the 11\,Mpc distance-limited sample of nearby galaxies \citep[11HUGS, LVL:][]{kennicutt08},
the starburst sample of \citet{engelbracht08}, 
and the two BCD samples studied by \citet{fumagalli10}, and \citet{hunt10}.
We added to these seven galaxy samples at higher redshifts:
the ``Green Pea'' compact galaxies identified by the Galaxy Zoo team at
$z\sim0.1-0.3$ \citep{cardamone09},
the LCG sample at $z\sim0.1-0.6$ selected by \citet{izotov11},
LBGs at $z\sim1$ \citep{shapley05a}, $z\sim2$ \citep{shapley04,erb06}, and
$z\sim3$ \citep{maiolino08,mannucci09}.
When stellar masses were not available in the literature,
we determined stellar masses from IRAC photometry as discussed in Sect. \ref{sec:masses}.
Altogether, we consider in the analysis 1070 galaxies from $z\sim0$ to $z\sim3$.
Except for the high-redshift samples \citep{shapley04,shapley05a,erb06,maiolino08,mannucci09}
and some of the LVL galaxies \citep{marble10},
metallicities for all galaxies
are determined by the ``direct'' (electron temperature) method \citep[e.g.,][]{izotov07}
since some of the samples are defined by requiring detections of \oiii$\lambda$4363.  
We adopt the central abundance of each galaxy, and thus
abundance gradients, if present, are not taken into account.
All stellar masses and SFRs are reported to the \citet{chabrier03} scale (see below for specific
instances).
Each sample is described in some detail in the following. 

\subsection{The dIrr sample } \label{sec:dIrr}

\citet{lee06} studied 25 dwarf galaxies within 5\,Mpc distance
that were observed at 4.5\,\micron\ with \spit/IRAC.
The advantage of using 4.5\,\micron\ to measure stellar masses
is that the stellar mass-to-light (M/L)
ratio does not depend strongly on age or metallicity, and dust
extinction is negligible.
\citet{lee06} derived M/Ls for their sample using the 4.5\,\micron\
luminosity, and the observed $B-$[4.5] color, transforming 
the 4.5\,\micron\ (Vega) magnitude, [4.5], to $K$ assuming
$K-$[4.5]\,=\,$+$0.2.
Sub-solar metallicity models from \citet{bell01} were then used to
derive M/L($K$) as a linear function of $B-K$; the solar
absolute magnitude at 4.5\,\micron\ was taken to be $\simeq+3.3$,
which implies that $K-M$ color is roughly 0.
All stellar masses are scaled to the
\citet{chabrier03} Initial Mass Function (IMF). 
Oxygen abundances for 21 galaxies are derived with the direct
method when electron temperatures are available from \oiii\,$\lambda$ 4363
measurements,
or the \citet{mcgaugh91} bright-line calibration for three additional objects.

SFRs are taken from \citet{woo08} and \citet{lee09},
based on extinction-corrected \ha\ luminosities using
the \citet{kennicutt98} conversion factor.
For 5 galaxies these were unavailable, so we adopted
the SFR values inferred from the star-formation histories
given by \citet{weisz11}.

A linear fit to the MZR for this nearby dwarf
sample gives a small scatter of 0.12\,dex,
over a wide range of stellar masses and metallicities:
$\sim$10$^6$ to 10$^9$\,\msun\ and 7.35 (Leo\,A) to 8.3 in O/H \citep{lee06}. 
The SDSS MZR has a comparable scatter, but does not extend either to low
stellar masses ($\ga10^{8.5}$\,\msun) or to low metallicities
(\logoh$\ga$8.3) \citep{tremonti04}.

\subsection{The 11HUGS/LVL sample} \label{sec:lvl}

\citet{kennicutt08} defined a distance-limited sample of 261 
late-type galaxies
(in the ``primary sample'') within 11\,Mpc, at high Galactic latitude
($|b|\geq20^\circ$), and with Hubble type $T\ge0$.
They also imposed a magnitude limit, $B<$15.
The 11\,Mpc Ultraviolet and \ha\ Survey (11HUGS) and the \spit\
Local Volume Legacy project (LVL) subsequently obtained {\it Galaxy Evolution Explorer}
(GALEX) imaging,
\ha\ photometry, and infrared (IR) imaging for the sample.
Because the LVL is a fair sampling of the Local Universe, it is dominated
by dwarf galaxies (see below).

The stellar masses have been derived by us from the 4.5\,\micron\
luminosities using the photometry given by \citet{dale09}. 
We have adopted the formulation of \citet{lee06}, but take 
the contamination by nebular and hot dust emission into account
as described in Sect. \ref{sec:masses}.
Stellar masses have also been estimated for the 11HUGS/LVL sample
by \citet{bothwell09} from the $B-V$ color and $B$-band luminosities,
using the formulation of \citet{bell01} for the dependence of M/L
on color\footnote{Because $B-V$ observations were only available
for roughly half the sample,
they assumed a standard $B-V$ color for the remaining galaxies.}.
Their masses range from 10$^7$ to $\ga10^{11}$\,\msun, similar to 
our values. 

Oxygen abundances are available for 129 of the 11HUGS/LVL galaxies, 
and are tabulated in \citet{marble10}e.
The values are from either \oiii\,$\lambda$4363, or
empirical strong-line methods, but have been analyzed by \citet{marble10}
and show no systematic deviations such as discussed by \citet{kewley08}.
The SFRs have been taken from \citet{lee09}, using the values derived
either from extinction-corrected \ha\ or UV luminosities, whichever are larger.
We considered both SFR values and chose the largest one
because \ha\ may not be a robust tracer of SFR at low SFRs
\citep[e.g.,][]{melena09,boselli09,lee11},
and we wanted to ensure an aggressive correction for nebular emission 
(see Sect.~\ref{sec:masses}).
There are a total of 113 11HUGS/LVL galaxies with all three quantities
(\mstar, O/H, and SFR) available.

The 11HUGS/LVL sample is dominated by dwarf galaxies; 90 (78\%) of 
these 113 LVL galaxies
have Hubble types $T\geq7$ \citep{lee09}, corresponding to Sd or later, and
79 (68\%) with $T\geq8$, later than but including Sdm.
As discussed by \citet{lee09}, there are a few dwarf spheroidals with
\logoh$\leq8.4$; 3 of these remain in our LVL subset.
SFRs generally range from $3\times10^{-4}$\,\msunyr\ to $\sim7$\,\msunyr;
NGC\,784 is a metal-poor (\logoh=7.9) outlier with 23\,\msunyr. 
Oxygen abundances run from 7.2 (UGC\,5340) to 9.3.
There are some weak active galactic nuclei in the LVL (e.g., NGC\,3726, NGC\,4826),
but they are all metal rich and we have eliminated them from the analysis.

\subsection{The BCDs} \label{sec:bcds}

The dwarf galaxies most likely to be metal poor with high SFRs are the BCDs.
These galaxies were defined originally on the basis of blue colors
and compact morphology \citep{thuan81}, but 
subsequently ``compact'' was superseded
by high (stellar) surface density \citep{gildepaz03}.
The parent samples from which most BCDs are drawn are generally
objective prism surveys (e.g.,
Markarian, SBS, UM, Tololo, etc.) which tend to preferentially
select nearby galaxies with high-equivalent-width emission lines.

Our BCD sample is the union of the starbursts studied
by \citet{engelbracht08},
and the galaxies investigated by \citet{fumagalli10} and \citet{hunt10}.
As described below, there are a total of 
89 BCDs in the combined samples (excluding 5 galaxies already in the LVL).
Not all the galaxies in the Engelbracht sample are BCDs or dwarf galaxies;
it was originally designed as a ``starburst sample'' with a wide range
of metallicities.
Hence, it adds high-mass star-forming galaxies to the mix.

Included here are also the two prototypical ``active'' and ``passive''
BCDs, \sbs\ and \izw\ \citep{hirashita04}.
Although these two galaxies have very similar metallicities, \logoh$\sim$7.2,
their SFRs differ by a factor of 10 \citep[1\,\msunyr\ vs. 0.1\,\msunyr,][]{hunt01,hunt05}.
We use the masses given by \citet{fumagalli10}, which agree well with
the masses we obtain here from the 4.5\,\micron\ luminosity 
(see Sect.~\ref{sec:masses}).

The distinction between a BCD and an "active" dIrr is not clear cut, and may
partially be a question of distance and the sensitivity with which extended
emission can be observed \citep{tolstoy09}.
In fact, not all the dwarf galaxies in the \citet{engelbracht08} sample are 
bona-fide (high surface brightness, compact) BCDs. 
There are two duplications with the LVL sample (UGC\,4483 and UGCA\,292);
in both cases, we use the LVL determination of O/H, SFR, and stellar masses.
On the other hand, some LVL galaxies are considered BCDs, e.g., NGC\,5253, NGC\,1140.
There are also some duplications among the BCDs; Mrk\,209 appears in both
the \citet{fumagalli10} and the \citet{hunt10} samples and
II\,Zw\,40, NGC\,2537, NGC\,5253, UM\,462, UM\,462 are in both
\citet{engelbracht08} and \citet{fumagalli10}.
In all cases, we use the \citet{fumagalli10} values for O/H and SFR.

For the BCD samples,
as for the 11HUGS/LVL sample, we derive stellar masses from the 4.5\,\micron\
luminosity, taking nebular and hot-dust contamination into account 
(see Sect. \ref{sec:masses}).
Oxygen abundances are taken from \citet{izotov07}, or from the
compilation by \citet{engelbracht08}. 
Virtually all of these are derived using \oiii\,$\lambda$4363,
so should be relatively robust in the face of systematic uncertainties
\citep[e.g.,][]{kewley08}.

SFRs were calculated from extinction-corrected \ha\ luminosities, or
alternatively, when the former were unavailable,
from  IR luminosities \lir\ according to the formulation given by \citet{draine07}
and the conversion factor of \citet{kennicutt98}.
The IR photometry came from \citet{engelbracht08} or from Hunt et al. (2012, in preparation).
When both \ha\ and IR photometry were available, we took the largest value of SFR;
in most cases (but not all), the \ha-derived SFRs are largest.

There are several weak active galactic nuclei (AGN) with super-solar
abundances in the Engelbracht et al. sample,
and we eliminated them from further consideration.
The stellar masses would be uncertain because of possible hot dust from the AGN,
and the abundances may also be incorrect because of AGN contamination of the
line emission.

There are 89 galaxies in the combined BCD sample.
The lowest abundance in this sample is \logoh\,=\,7.1 for \sbs W. 
In the metallicity range we are considering,
stellar masses (see below) range from $2\times10^5$\,\msun\ to $1.6\times 10^{11}$\,\msun,
and
SFRs range from $10^{-4}$\,\msunyr\ (DDO\,187)
to 74\,\msunyr\ (SHOC\,391, roughly 3 times higher than the 
largest LVL SFR).

\subsection{The Green Peas and the LCGs} \label{sec:peas}

Recently, \citet{cardamone09} identified an unusual class of galaxies
noted for their compact, green appearance on the SDSS composite images, 
the so-called ``green pea galaxies''.
These galaxies appear green 
because of their very strong \oiii $\lambda$5007 optical emission line redshifted
into the SDSS $r$ band. 
The Green Peas have redshifts of $0.11 < z < 0.36$, and are low mass
($10^{8.5} - 10^{10}$\,\msun) with high SFRs ($\sim 10$\,\msunyr).
Their metallicities were reexamined by \citet{amorin10} and by \citet{izotov11}
and found to be somewhat metal poor, \logoh $\sim$ 8.0.
They are offset by $\sim$0.3\,dex in the MZR from the galaxies in the Local Universe
\citep[e.g.,][]{tremonti04}, and are thought to be local analogues of 
more distant ultraviolet-luminous galaxies such as LBGs \citep{cardamone09}.

\citet{izotov11} selected a larger sample of similar galaxies, the LCGs, 
by introducing a threshold in (extinction-corrected) \hb\ luminosity ($3\times10^{40}$\,\ergs),
requiring a high \hb\ equivalent width ($>$50\,\AA), 
and including only those objects with a well-detected \oiii $\lambda$4363 emission line
in order to obtain accurate abundances using the direct method \citep[in contrast to][who
used the less accurate empirical strong-line method]{cardamone09}.
Moreover, a compact appearance was required on the SDSS images, and galaxies with
evidence of AGN spectral features were excluded.
These criteria resulted in a LCG sample of 803 galaxies,
with redshifts ranging from 0.1 to 0.6 (although most of them have $0.1<z<0.3$), 
and metallicities from \logoh\ from 7.5 to 8.4,
with a median value of $\sim$8.1.
SFRs are derived from extinction-corrected
\ha\ luminosities and range from 0.7 to $\sim60$\,\msunyr.
\citet{izotov11} compared these values with SFRs inferred from FUV luminosities from GALEX,
and found good agreement. 

The LCGs and the 
Green Peas provide excellent examples of low-metallicity starbursts. 
They are at sufficiently low redshifts ($z\sim0.1-0.3$) that they should not present
significant evolutionary differences relative to the less distant metal-poor starbursts
described in the previous sections.
On the other hand, their properties are sufficiently extreme in terms of low oxygen
abundance and high SFR that they can be good local counterparts for high-redshift LBGs,
although they are much fainter \citep[$\sim 3$\,mag,][]{izotov11}, and less
massive (see below).
The SFR maximum value, $\sim60$\,\msunyr\ of the LCGs and Green Peas is virtually identical to the
maximum values ($\sim60-66$\,\msunyr)
in the LBG $z\sim2$ and $z\sim3$ samples described by \citet{steidel04} 
and \citet{steidel03}.
However, the minimum SFR in the high-redshift LBG samples is 
$\sim3-6$\,\msunyr, while in the LCGs is lower
(by definition, because of the threshold in \ha\ luminosity), 0.7\,\msunyr.

Here we include in our analysis the \citet{izotov11} LCG sample of 803 galaxies, 
together with the subset
of \citet{cardamone09} Green Pea galaxies (66) for which \citet{izotov11} remeasured the 
masses and metal abundances (see their Table 2).
\citet{izotov11} derived stellar masses by fitting the SDSS spectrum from 
0.39 to 0.92\,\micron, and approximating the star-formation history with two
episodes: a recent short burst and a previous continuous star-formation event
responsible for the evolved stars.  
They used a Salpeter IMF \citep{salpeter55}, and we have divided by a correction factor of 1.8
\citep[see][]{lee06,erb06,pozzetti07} to
bring the masses and SFRs to the \citet{chabrier03} scale.

\subsection{The LBGs} \label{sec:lbgs}

We use LBGs as a natural high-redshift galaxy population to compare with the local
metal-poor starbursts.
12 LBGs in the DEEP2 sample at $z\sim1$ were characterized by \citet{shapley05a}.
The weakness of the \oiii$\lambda$4363 line precluded direct methods for measuring
nebular oxygen abundance, so they used strong-line techniques (N2, O3N2).
Stellar masses were calculated by fitting the $BRIK_s$ 
broadband spectral energy distributions (SEDs) with synthetic stellar populations
from \citet{bruzual03} and exponentially declining SF histories.
They derived SFRs from \ha\ luminosities with the conversion of \citet{kennicutt98}.

\citet{erb06} studied a sample of LBGs at $z\sim2$, and derived stellar masses
by fitting the broadband SEDs using $U_nGRJK$,
and, when available, \spit/IRAC bands to standard spectral synthesis models
from \citet{bruzual03}.
SFRs were calculated from extinction-corrected \ha\ luminosities, and  
metallicities were established using the strong-line method because of the
weakness of \oii $\lambda$4363 in these high-redshift objects.
They arranged their sample in 6 bins of stellar mass, in order to ensure
robust trends with mass and metallicity.
\citet{erb06} use the \citet{chabrier03} IMF, so we have applied no correction
to the masses or the SFRs.
Another sample of 7 LBGs at $z\sim2$ was studied by \citet{shapley04}.
They fit broadband SEDs ($U_nGRK_s$) to \citet{bruzual03} populations with
a Salpeter IMF, so we have divided by a factor of 1.8 to correct them to
the Chabrier IMF.
As above, oxygen abundances were calculated from strong-line methods,
and SFRs from \ha\ luminosities.

\citet[][AMAZE]{maiolino08} and \citet[][LSD]{mannucci09} studied 18
similarly selected LBGs at $z\sim3$ but also required that at least two 
of the \spit/IRAC bands be present in order to obtain more reliable stellar masses.
\citet{mannucci09} fit $UGRJK+3.6,4.5\mu$m broadband SEDs with models from 
HYPERZ-MASS \citep{pozzetti07};
HYPERZ-MASS considers a Chabrier IMF, so we have not introduced any correction
factors for their objects. 
In \citet{maiolino08}, 
stellar masses are derived by fitting the broadband SED (with the same
photometry as above) with a series of standard spectral synthesis models 
as described by \citet{fontana06}.
They used a Salpeter IMF, so -as above- we have 
divided stellar masses and SFRs by a correction factor of 1.8 to be consistent 
with our adopted Chabrier IMF. 
Both papers estimate metallicities by comparing three independent
strong-line calibrations of the direct method. 

\subsection{Stellar masses from IRAC 4.5 $\mu$m} \label{sec:masses}

For the dwarf galaxy samples in the Local Universe, we
calculated the stellar masses from \spit/IRAC 4.5\,\micron\ luminosities,
using a variation of the method described in \citet{lee06}.
Near-infrared bands ($\lambda\geq$1\,\micron) in general, and
IRAC bands in particular (because of their availability and sensitivity)
are particularly well suited for estimating stellar mass.
At these wavelengths, the stellar M/L ratio is quite insensitive to
age and metallicity \citep[e.g.,]{gavazzi93,jun08}, and dust extinction is negligible.
However, nebular emission in metal-poor starbursts can be very significant,
both in the continuum and in hydrogen recombination and other lines
\citep{reines10,atek11}.
This can have a potentially strong impact on the 4.5\,\micron\ photometry \citep{smith09},
and thus on the stellar masses derived from it.
Hence, before applying the formalism of \citet{lee06}, we have estimated the
strength of \bra\ emission and the nebular continuum in the 4.5\,\micron\ IRAC
band, and subtracted it from the observed flux.

To estimate the nebular component at 4.5\,\micron, we first calculated
the SFR, either from 
\ha\ luminosities when available,
or from \lir\ \citep{draine07,kennicutt98} if not.
Then,
assuming the Case~B recombination coefficients given by \citet{osterbrock06}, 
we used the observed \ha\ flux \citep[or that inferred from the
conversion of SFR to \ha :][]{kennicutt98,lee09}
to calculate \bra\ flux.
The nebular continuum in the 4.5\,\micron\ IRAC band was inferred
by interpolating to IRAC wavelengths
the volume emission coefficients given by \citet{osterbrock06}.

At low metallicities, \te\ tends to be rather high, $\simgt$10000\,K 
and can even exceed 20000\,K \citep[e.g.,][]{guseva06,izotov07}. 
Hence, all calculations were performed assuming two different electron temperatures, \te,
10000\,K and 20000\,K.
Because we were unable to uniformly derive \te\ for all the galaxies in our
dwarf samples, we simply took the average of these two values;
this results in an intrinsic uncertainty of $\sim$10\% for the gas contribution.

Once the average predicted nebular emission, both \bra\ and continuum, was subtracted from
the observed 4.5\,\micron\ photometry, we applied the method developed by
\citet{lee06}.
This involves a M/L ratio based on the $B-K$ color, 
and a correction to the 4.5\,\micron\ band to approximate $K$,
as described in Sect. \ref{sec:dIrr}.

Because hot-dust emission in low-metallicity starbursts
can also be significant at or longward of 2\,\micron\ \citep{hunt01,hunt02},
we also attempted to infer the contamination at IRAC wavelengths from hot dust.
This was possible when 3.6\,\micron\ fluxes were available\footnote{This is
true only for the LVL and Engelbracht samples.}.
After subtracting the nebular emission from the 3.6\,\micron\ fluxes,
we then estimated the stellar emission at 4.5\,\micron\ using the formulation
given by \citet{helou04}, i.e., 0.596 times the 3.6\,\micron\ emission. 
The excess 4.5\,\micron\ emission, after correcting for nebular contamination
and subtracting the stellar component, should be largely due to hot dust.

The amount of contamination in the nearby dwarf galaxies, 
both from nebular emission and from hot dust, can be significant.
In some sources (notably HS\,0822$+$3542, I\,Zw\,18, II\,Zw\,40,
Mrk\,209, Mrk\,1315, SHOC\,391, Tol\,65), 
the contamination at 4.5\,\micron\ from \bra\ alone can exceed 10\%.
Nebular contamination at 4.5\,\micron\ can be as high 25-30\%,
although the median value among the dwarf samples is $\sim$10\%.
The hot dust emission at 4.5\,\micron\ can also be a relatively large
fraction of the total:
in NGC\,5253, \sbs, Haro\,11, Pox\,4, the hot dust contamination 
of the IRAC 4.5\,\micron\ band is $\ga$40\%.

There are 7 BCDs studied by \citet{fumagalli10}
that have stellar masses derived from combinations of optical and
near-infrared colors that are in common with our sample.
In general, the masses agree to with $\sim$30\% (0.1\,dex).

\subsection{Caveats}

In the end, our sample consists of 16 dIrrs \citep{lee06}, 
113 LVL galaxies \citep{kennicutt08},
89 local starbursts and BCDs \citep{engelbracht08,fumagalli10,hunt10},
803 Green Peas and LCGs \citep{izotov11},
and 43 LBGs at $z\sim1-3$ \citep{erb06,shapley04,shapley05a,maiolino08,mannucci09}.
This combined sample of $\sim$1070 galaxies has stellar masses
and metallicities derived with heterogeneous methods;
these could affect our analysis of scaling relations, but as we argue below,
do not.

The $z=0$ samples have masses based on the 4.5\,\micron\ luminosity
which minimizes variations in the M/L ratio \citep{jun08}.
We have also removed nebular and recombination line emission
and applied the $B-K$ color correction to the M/L ratio as advocated
by \citet{lee06}.
Hence, these masses should be sufficiently reliable to compare with
those derived from the more sophisticated SED fitting methods applied 
to the higher-redshift samples.
It is also true that our sample spans $10^5$ in stellar masses, so
any systematics should be overcome by sheer dynamic range.

As described in Sect.~\ref{sec:lvl},
for the LVL/11HUGS galaxies we used the largest of the UV or \ha-derived SFRs.
Relative to SFRs inferred from the FUV continuum,
SFRs calculated with \ha\ can be underestimated at low SFRs.
Although the differences could be due to the uncertain extinction corrections \citep{boselli09},
they could also result from effects in stochasticity in the number of
massive stars in small clusters.
In any case, they should be directly comparable to \ha-inferred values
for higher SFRs in other samples \citep[e.g.,][]{lee09},
and should not affect our main conclusions about metal-poor starbursts.

The metallicities are potentially more problematic since most of the
local galaxies and all of the LCGs/Green Peas have O/H derived from
the direct electronic temperature method.
On the other hand, the high-$z$ samples rely on various strong-line calibrations, so
there may be some discrepancy between the two sets of abundances.
Nevertheless, according to \citet{kewley08}, the largest deviations
(up to $\sim$0.7\,dex) 
occur at the highest metallicities and for the photoionization technique.
Typical discrepancies between the $T_e$ and the strong-line methods
should be $\la$0.15\,dex, at least for the relative comparison.
Since our sample is dominated by sub-solar metallicities, and because
the range in metallicities is so large ($\sim\,\times$100), we do not
expect any problems with the metallicity determinations to exert
a strong influence on the MZR for our sample.

\section{Low-metallicity starbursts} \label{sec:lms}

\begin{figure}
\includegraphics[width=\linewidth,bb=18 144 592 650]{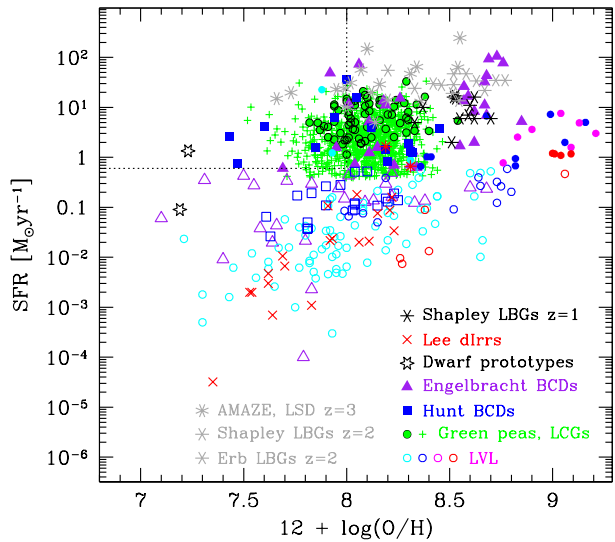} 
\caption{SFR plotted against nebular oxygen abundance.
The LVL (11HUGS) galaxies are shown as small open or filled circles
(filled when SFR$\geq$0.6\,\msunyr),
with different colors corresponding to Hubble type $T$ as in
\citet{lee09}: $T \geq$ 8 cyan, $5\leq T<8$ blue,  $3\leq T<5$ magenta,
and $T<3$ red.
Red $\times$ corresponds to the dIrr sample;
BCDs are shown as (blue) squares (Fumagalli$+$Hunt samples) and 
(purple) triangles (Engelbracht);
Green Peas are given by small (green) filled circles;
LCGs by $+$;
LBGs at $z\sim1$ (Shapley), $z\sim2$ (Erb, Shapley) are shown as 6-pronged asterisks, 
and at $z\sim3$ (Maiolino, Mannucci) as 8-pronged asterisks. 
6-sided open stars show the two dwarf ``prototypes'', \sbs\ and \izw. 
Solid symbols show those BCDs with SFR$\geq$0.6\,\msunyr.
The region in the upper left quadrant delineated by dotted lines
corresponds to what we loosely define as a ``low-metallicity starburst" (see text):
SFR$\geq$0.6\,\msunyr\ and \logoh$\leq$8.0.
}
\label{fig:sfr_oh}
\end{figure}

\begin{figure}
\includegraphics[width=\linewidth,bb=18 144 592 650]{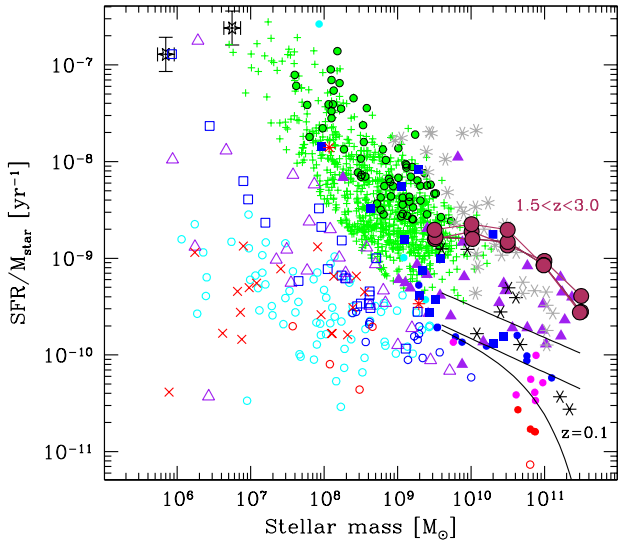} 
\caption{sSFR plotted against stellar mass, \mstar\ (star-formation ``Main Sequence'').
Symbols are as in Fig.~\ref{fig:sfr_oh}.
The black curve and line correspond to the Local-Universe \citep[$z=0.1$,][]{salim07};
the curve reproduces the best-fit Schecter function,
and the line the best fit to ``pure'' (no AGN) star-forming galaxies.
The black line with the higher normalization gives
the intermediate-redshift trend \citep[$0.2<z<0.7$,][]{noeske07}; and the bordeaux
curves with round data points indicate the high-redshift relations 
\citep[$1.5<z<3.0$,][]{bauer11}.
}
\label{fig:ms}
\end{figure}

\begin{figure}
\includegraphics[width=\linewidth,bb=18 144 592 650]{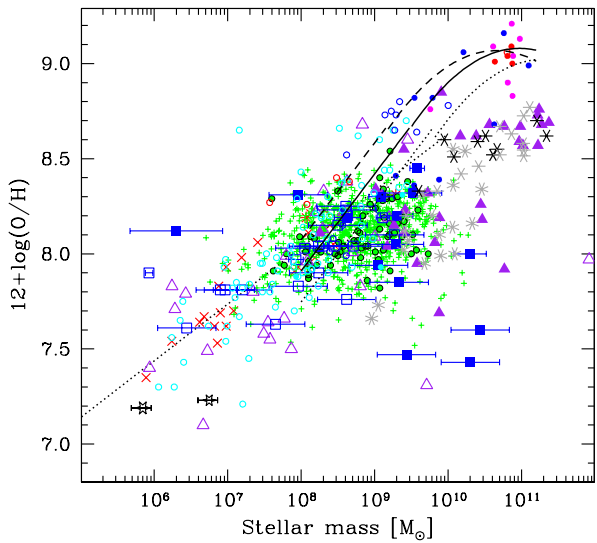} 
\caption{Nebular oxygen abundance, \logoh, plotted against stellar mass, \mstar\ (MZR, FMR).
Symbols are as in Fig.~\ref{fig:sfr_oh}.
The black dotted line gives the linear relation found by \citet{lee06},
and the curves correspond to the adjustments for SFR as advocated by \citet{mannucci11}.
SFRs consistent with the ranges considered by \citet{mannucci10,mannucci11}
are shown (0.1\,\msunyr, dashed line to the left; 1\,\msunyr, solid line;
10\,\msunyr, dotted line to the right). 
}
\label{fig:mzr}
\end{figure}

Figures \ref{fig:sfr_oh}, \ref{fig:ms}, and \ref{fig:mzr}
show the trends of SFR, oxygen abundance, and stellar mass
for all samples.
Here we focus on the outliers
from the general relations and identify a ``low-metallicity starburst".

Generally SFR increases with metallicity, although the correlation
is far from perfect. 
Inspection of Fig.~\ref{fig:sfr_oh}, which plots SFR against \logoh,
shows that in two of the local dwarf samples (LVL, dIrr) there is a clear
increase of SFR with metallicity.
The trend gets rather noisy for the BCD samples, and disappears completely
for the LCG$+$Green Peas.
Nevertheless, the statistical significance of the correlation of SFR
with O/H over all the samples (1070 data points) is quite high, $\ga 10\,\sigma$. 
In fact, two regions tend to be underpopulated:
the lower right quadrant with high abundances and low SFRs,
and the upper left quadrant with low O/H and high SFRs.
The galaxies that occupy this latter region are what we will call
``low-metallicity starbursts''.

Figure \ref{fig:ms} shows the ``Star Formation Main Sequence'' or SFMS. 
In addition to the data, three trends are shown as black curves
extending only as far as the parameter space in the original samples they were derived from.
The SFMS derived by \citet{salim07} for the Local Universe at $z\sim0.1$
(both the linear relation and the ``Schechter-like'' curve) appears
at lower sSFRs;
the one proposed by \citet{noeske07} for galaxies at redshifts between 0.2 and 0.7
is roughly parallel to the previous one but with a larger offset;
and the SFMS derived by \citet{bauer11} for a mass-selected
sample between redshifts 1.5 and 3.0 has an even greater offset. 
The figure clearly shows that most of the AMAZE and LSD samples at $z\sim3$ follow
the Bauer et al. relation fairly well.
However, the high-$z$ trend is also reflected in the LCGs and the Green Peas at $z\sim0.3$, 
and several of the BCDs at $z\sim0$.
Most of the BCDs follow the Local Universe SFMS by \citet{salim07}, but some
galaxies at low stellar masses have excess SFR relative to the SFMS.
At stellar masses \mstar$\sim10^8$\,\msun, the SFR of these outliers is $\ga0.6$\,\msunyr.
This motivates our choice of SFR\,=\,0.6\,\msunyr\
as a SFR threshold for starbursts at low metallicity.
Such a low-mass galaxy would have an sSFR of $\sim 6\times10^{-9}$\,yr$^{-1}$,
well outside the range probed by \citet{salim07}, but
roughly consistent with
an extrapolation to lower masses of the ``starburst sequence'' they define.

The MZR for our data is shown in Figure \ref{fig:mzr}.
The FMR proposed by \citet{mannucci10} is also shown as a set of black curves;
these curves are the predictions, based on the SDSS, of how SFR alters
the MZR. The dashed curve corresponds to SFR\,=\,0.1\,\msunyr, 
i.e., where a galaxy with this SFR would lie relative to the MZR.
The solid curve gives a correction for SFR\,=\,1\,\msunyr, and the
dotted one SFR\,=\,10\,\msunyr;
the curves span roughly the ranges in parameters of the \citet{mannucci10} sample.
The FMR was subsequently extended to lower masses by \citet{mannucci11}
(shown here as the linear extensions toward lower masses slightly detached 
from the main curved FMR).
The linear MZR for low masses found by \citet{lee06} is also shown, but
the slope with \mstar\ is flatter ($\sim$0.3) than the slope found by
\citet[][$\sim$0.5]{mannucci11}.

Most of the AMAZE and LSD galaxies at $z\sim3$ follow neither the MZR
nor the FMR, and this was already noticed by \citet{mannucci10}. 
However, some of the low-metallicity dwarf galaxies with SFR$\ga$0.6\,\msunyr\
are also outliers of both the MZR and the FMR,
although galaxies with lower SFR follow the MZR and/or FMR quite well. 
The degree of deviation from the main trends increases
for metal abundances at or below \logoh$\sim$8.0.
Some metal-poor BCDs and high-$z$ LBGs in the AMAZE$+$LSD samples
deviate from the MZR and FMR by factors of 100 or more in stellar mass.
Hence, we adopt this abundance \logoh\,=\,8.0 as a ``fiducial location'' 
in the MZR below which we can distinguish potential outliers; 
here we are interested in low-metallicity starbursts, and consequently concentrate
on this metal-poor abundance regime and on high SFRs.
The region delineated in Fig.~\ref{fig:sfr_oh} is occupied by these
metal-poor starbursts, independently of their redshift.
If a galaxy has \logoh$\leq$8.0 and SFR$\geq$0.6\,\msunyr, we will call it
a ``low-metallicity starburst'' (LMS).

\subsection{Crude demographics} \label{sec:demographics}

The only galaxy sample in our database which can be considered complete in
any sense is the 11HUGS/LVL sample; it is approximately distance (thus volume) limited,
and a fair representation of the nearby galaxy population.
Of 113 galaxies in the LVL sample, only 2 (NGC\,784 and NGC\,4656) would be 
considered LMSs, a fraction of $\sim$ 2\%.
The fraction of low-metallicity starbursts in the BCD samples is 5 times higher,
$\sim$11\% (10 LMSs),
but these galaxies have already been selected from the general galaxy population
through the strength of their emission lines.

Among the LCGs and Green Peas, the fraction is higher still, but, again,
these galaxies have been selected for strong emission lines; 
176 galaxies out of 803 LCGs would be classified as a LMS, roughly 22\%. 
This is partly a result of a selection effect because of the limit on \hb\
luminosity, and thus SFR, imposed by \citet{izotov11}; {\it all} LCGs have
SFRs greater than our fiducial cutoff by definition.
At $z\sim3$,
there are 3 LMSs in the LSD sample of 9 and 2 LMSs of 9 AMAZE galaxies,
or $\sim$27\%;
these are small-number statistics, but roughly consistent with the LCGs
at a redshift 10 times smaller.

Outside the LVL and dIrr samples, virtually all the galaxies are
selected on the basis of high equivalent widths in various emission lines.
These samples have high LMS fractions, 
perhaps more similar to the LBGs at $z\sim3$ than either the LVL or dIrr samples.
This means that the differences in properties found for high-$z$ 
populations may result from differences in selection criteria,
relative to the local galaxy populations.
We may be observing locally a
somewhat rare class of low-metallicity starbursts that are however
similar to some high-$z$ galaxy populations such as LBGs.
In the next section, we put this proposition on a firmer footing
by showing the existence of a Fundamental Plane of O/H, SFR, and \mstar,
which apparently does not vary with redshift.



\section{The fundamental plane of metallicity, SFR, and stellar mass} \label{sec:fp}

Unlike the SDSS samples studied by other groups \citep{tremonti04,salim07,mannucci10,yates12},
our sample has been designed to probe low stellar masses, low metallicities,
and (relatively) high SFRs.
This makes it uniquely suited to exploring a much expanded parameter space,
and the resulting scaling relations of metallicity, \mstar, and SFR.
At high masses and metallicities, both the MZR and SFMS inflect and flatten.
However, for \mstar$\leq3\times10^{10}$\,\msun, roughly the ``turn-over mass''
\citep{tremonti04,wyder07}, the trends are very close to linear, independently
of redshift.
Hence, the observationally-defined variables (O/H, SFR, and \mstar) should
define a {\it plane} which, given the relatively large scatters in the SFMS, the
MZR, and the SFR-O/H relation, is not optimally viewed
\citep[see also][]{laralopez10}.

We can implicitly test the assumption that the relations among the observationally-defined variables
are invariant, regardless of redshift. 
Hereafter, for brevity,
we will call the ``observationally-defined variables'' {\it observables}, even
though they are {\it derived} from observations rather than being directly observed.
Like \citet{mannucci10}, we propose that
if the MZR and SFMS are
observed to ``evolve'' or change with redshift, then this evolution or
alteration must be due to 
different selection criteria for different galaxy
populations at different lookback times.
However, \citet{mannucci10} defined a ``surface'' by associating SFR with
stellar mass, and plotting this combination against metallicity;
they did not examine the mutual correlations among the three parameters. 
Metallicity, mass, and SFR are all correlated
(see Sect. \ref{sec:lms}), but
it is not clear which parameters are mainly responsible for the dispersion.

We have therefore performed a Principle Component Analysis (PCA)
on all galaxies with stellar masses $\leq3\times10^{10}$\,\msun\ (1019 data points).
A PCA diagonalizes the covariance matrix, thus giving the linear combinations of 
observables which define the orientations that minimize the covariance. These 
orientations are the eigenvectors and are, by definition, mutually orthogonal.
In the 3-space formed by the three observables, to form a plane
we would expect most of the variance to be contained in the first 
two eigenvectors; for the third, perpendicular, eigenvector the variance should be 
very small.

Our sample is particularly well suited for such an analysis; it spans
almost two orders of magnitude in metallicity (\logoh\,=\,7.1 to $\sim9$),
a factor of $\sim10^6$ in SFR ($\sim10^{-4} \leq$ SFR $\leq\, \sim10^2$\,\msunyr),
and a factor of $\sim10^5$ in stellar mass ($\sim10^6 \leq$ \mstar\ $\leq\, \sim10^{11}$\,\msun).
Other samples previously analyzed to find scaling relations cover much smaller
parameter ranges: typically less than a decade in metallicity (\logoh$\geq$8.4), 
a factor of $\sim$200 in SFR ($\sim$0.04$\la$ SFR$\la 6$\,\msunyr),
and roughly 2 orders of magnitude in stellar mass (\mstar$\ga 10^9$\,\msun)
\citep[e.g.,][]{tremonti04,salim07,mannucci10,yates12}. 
This implies that our results should represent the extremes of
SF processes, possibly closer to what is observed in the early universe than nearby.
To avoid excessively weighting the LCGs, which effectively dominate our
sample ($\sim$800/1000), we used a bootstrapping technique to randomly select
a fraction of LCGs and repeat the PCA many times on 
different samples of $\sim$350-1000 objects.

\begin{figure*}
\hbox{
\includegraphics[width=\linewidth,bb=18 144 592 600]{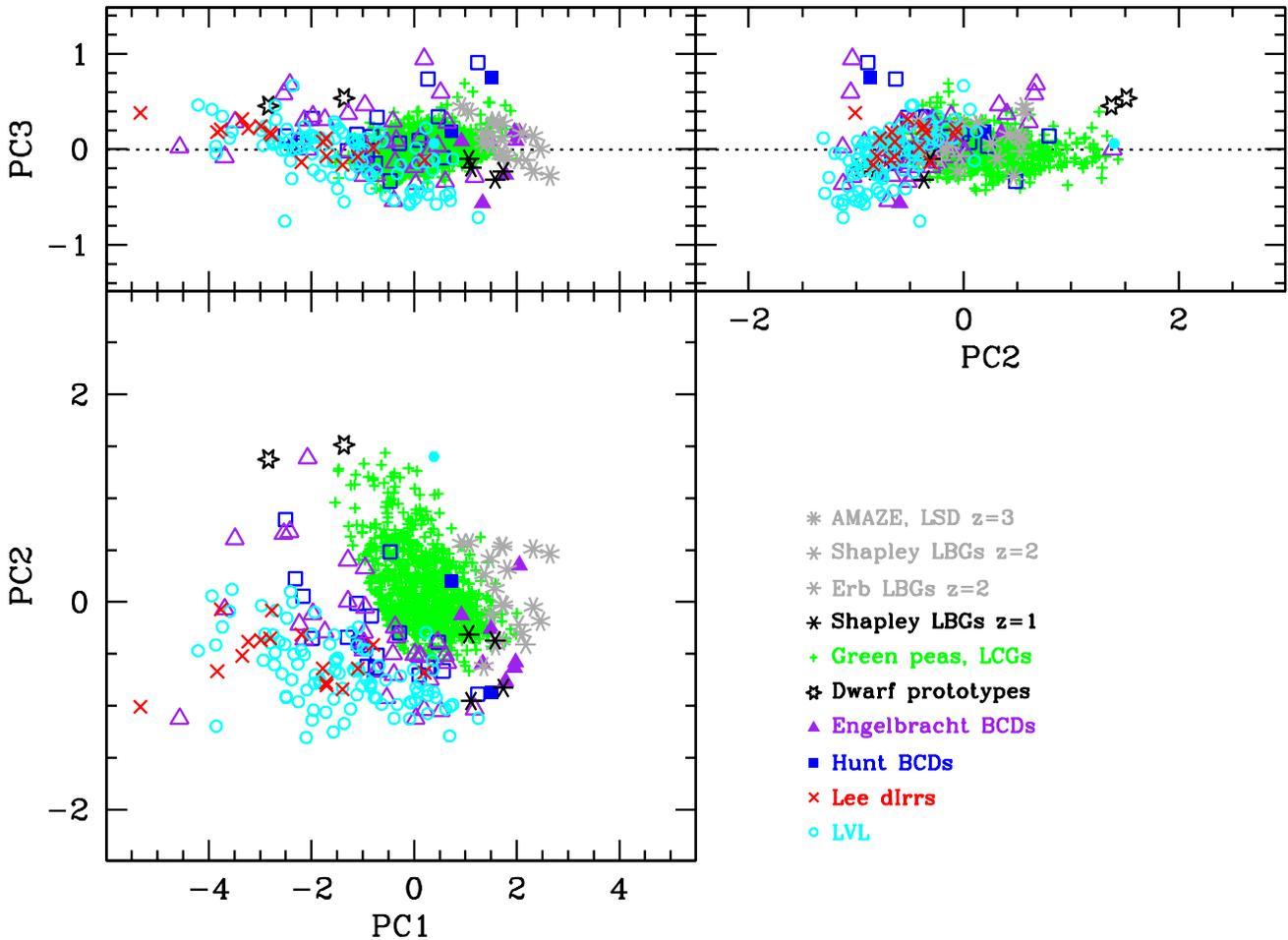} 
}
\caption{Different projections using the 3 PCs found by the PCA.
Galaxies are coded as shown in the lower right (empty) panel.
The top left and right panels show the orthogonal ``edge-on'' views of the plane; 
the bottom panel shows the plane face-on.
$x_1\,=\,$\logoh$-\langle$ \logoh $\rangle$,
$x_2\,=\,\log (SFR)-\langle \log(SFR) \rangle$ (\msunyr), and
$x_3\,=\,\log (M_{\rm star})-\langle \log(M_{\rm star}) \rangle$ (\msun).
$\langle$\logoh$\rangle$\,=\, 8.063;
$\langle\log(SFR)\rangle$\,=\, $-0.594$;
$\langle\log(M_{\rm star})\rangle$\,=\,8.476.
PC1 \, = \, 0.12\,$x_1$ + 0.75\,$x_2$ + 0.65\,$x_3$; 
PC2 \, = \, -0.31\,$x_1$ - 0.65\,$x_2$ - 0.69\,$x_3$;
PC3 \, = \, -0.94\,$x_1$ - 0.11\,$x_2$ + 0.31\,$x_3$.
}
\label{fig:pca}
\end{figure*}

\begin{figure*}
\hbox{
\includegraphics[width=0.5\linewidth,bb=18 144 592 718]{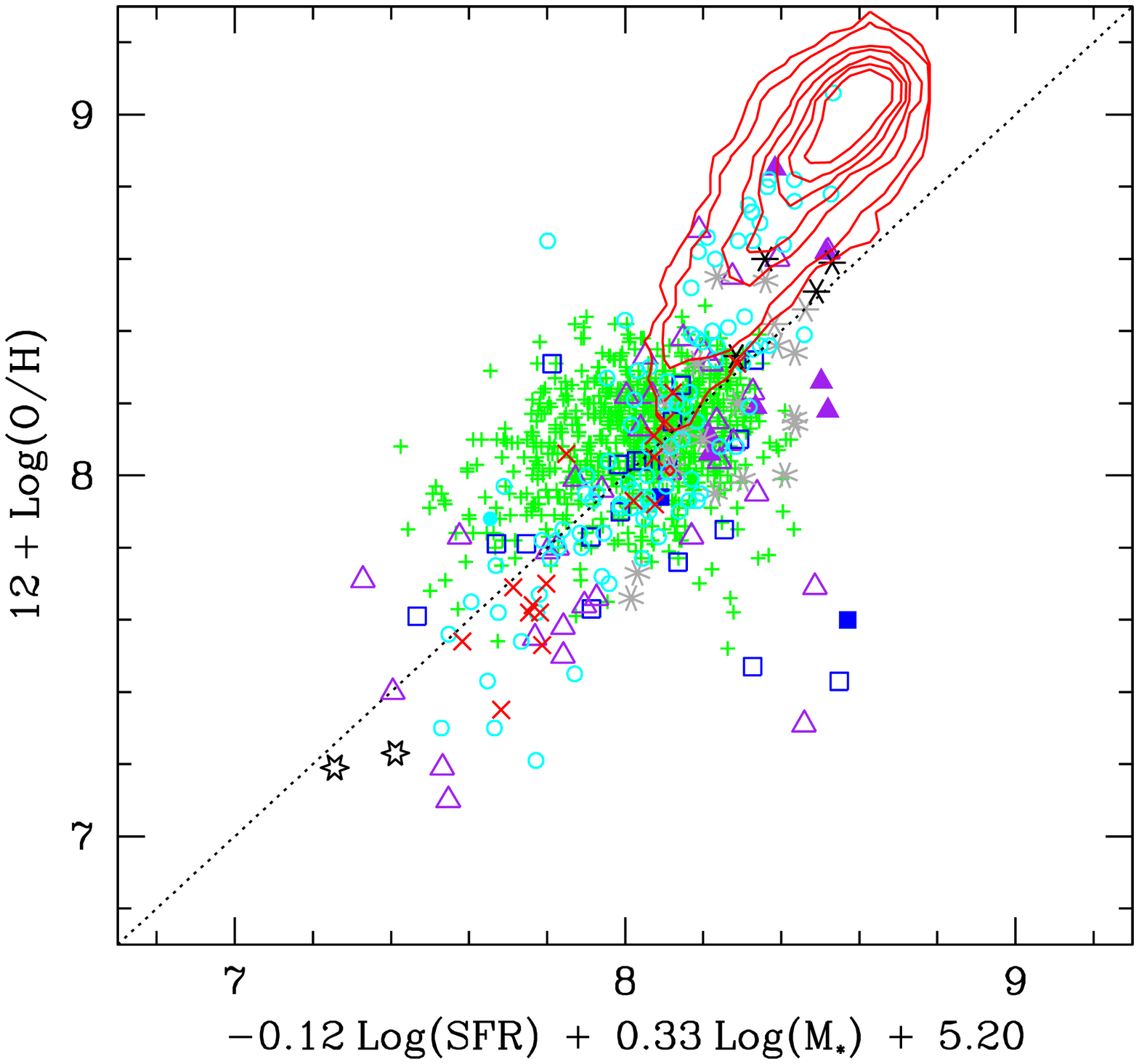} 
\includegraphics[width=0.5\linewidth,bb=18 144 592 718]{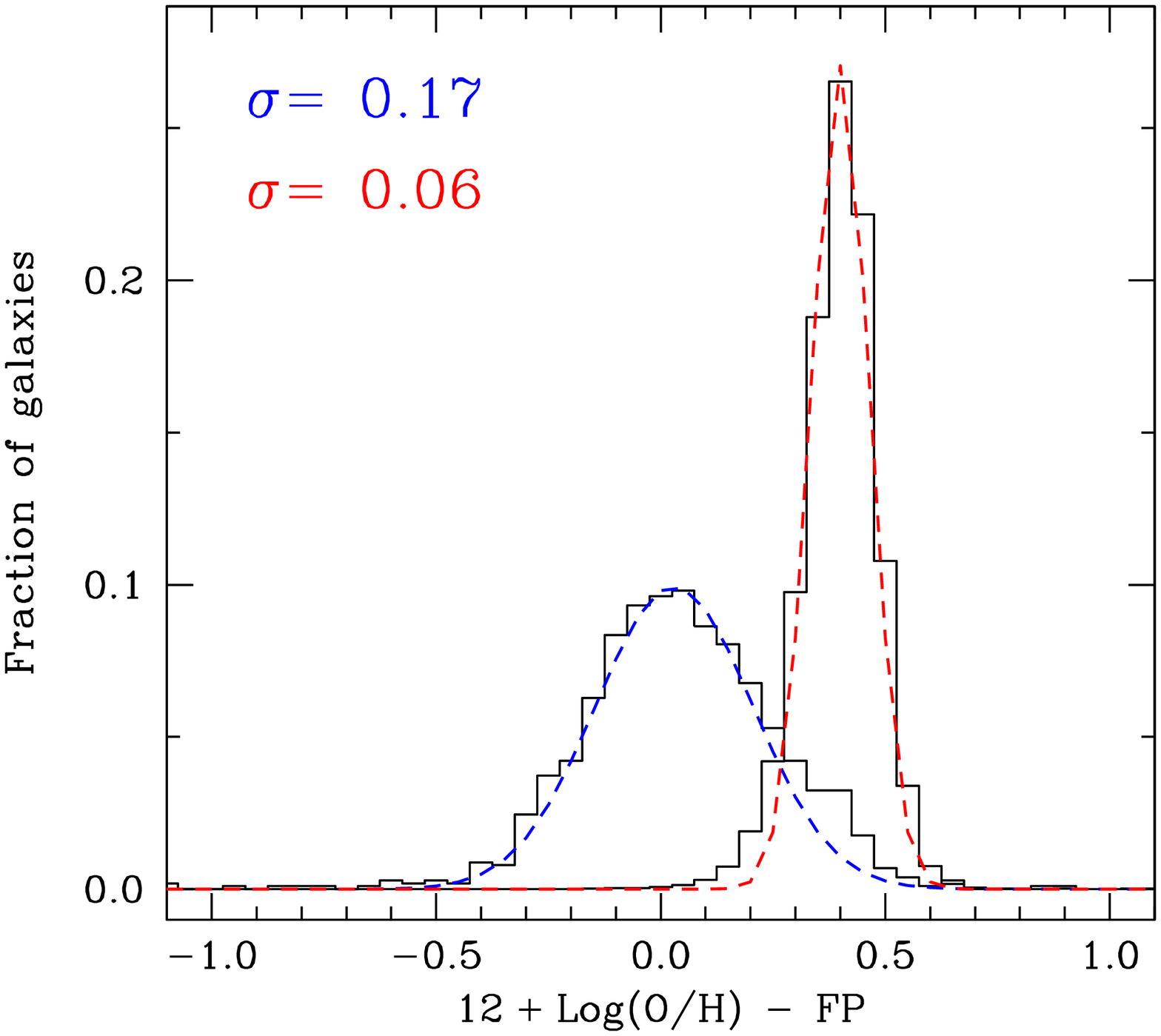} 
}
\caption{Left panel: the PC3 component shown as \logoh\ vs. the resulting 
linear combination of (log) SFR and \mstar\ obtained from the PCA on our sample.
Galaxies are coded as in Fig.~\ref{fig:pca}.
The FP applied to the FMR-SDSS sample is shown as red contours,
with the innermost contour corresponding to 60\% of the sample of 55\,082 galaxies,
and increasing to 70\%, 80\%, 90\%, 95\%, 99\%, with 99.5\% as the outermost contour.
The dotted line in the left panel indicates equality.
Right panel: the histogram of the residuals from the identity (dotted line)
relation shown in the left panel;
the blue Gaussian and broader distribution corresponds to our sample,
and the red Gaussian and narrower distribution to the FMR-SDSS galaxies.
The FMR-SDSS distribution is well fit by a Gaussian but offset by $+0.4$\,dex.
In both samples (and both panels), only galaxies with \mstar\ $< 10^{10.5}$\,\msun\ are shown.
}
\label{fig:fp}
\end{figure*}

The PCA shows that the set of three observables truly defines a plane;
$\sim$97-98\% of the total variance is contained in the first two eigenvectors.
80-85\% (depending on the bootstrapping) of the variance is contained
in the first eigenvector or Principle Component (PC), PC1;
12-17\% of the variance is found in the second PC, PC2;
and $\la$2-3\%  in PC3.
Hence, the 3-space defined by O/H, SFR, and \mstar\ is degenerate; 
only two independent parameters are necessary to 
describe the galaxies in these three dimensions.

The existence of a plane of O/H, SFR, and \mstar\ for star-forming galaxies
is not, however, a new result. 
\citet{laralopez10} found that a plane could describe a sample of 
$\sim$ 33\,000 SDSS galaxies but, rather than performing a PCA, they fitted
linear regressions to pairs of observables.
They also included a second-order polynomial for low-metallicity mass bins,
and concluded that a ``linear combination of the metallicity and SFR as a
function of stellar mass'' could describe their sample.
Thus, unlike \citet{mannucci10} who combined SFR and \mstar\ to predict \logoh,
\citet{laralopez10} combined SFR with \logoh\ to predict \mstar, and find
a dispersion $\sigma$ of 0.16 for the 33\,000
SDSS galaxies in their sample.

For our sample,
Fig. \ref{fig:pca} shows three different projections of the plane
defined by these three ``basis'' (eigen-) vectors that are different
linear combinations of the observed variables:
\begin{eqnarray}
PC1 & = & 0.12\,x_1 + 0.75\,x_2 + 0.65\,x_3 \\
PC2 & = & -0.31\,x_1 - 0.65\,x_2 - 0.69\,x_3 \\
PC3 & = & -0.94\,x_1 - 0.11\,x_2 + 0.31\,x_3
\end{eqnarray}
\noindent
The PCA first normalizes the variables to zero mean, so that
\begin{eqnarray}
x_1 & = & 12+\log(O/H) -\langle  12+\log(O/H) \rangle \nonumber \\
    & = & 12+\log(O/H)-8.063 \nonumber \\
x_2 & = & \log (SFR)-\langle \log(SFR) \rangle \nonumber \\
    & = & \log(SFR)+0.594 \ \ M_\odot\,{\rm yr}^{-1} \nonumber \\
x_3 & = & \log (M_{\rm star})-\langle \log(M_{\rm star}) \rangle \nonumber \\
    & = & \log M_{\rm star} -8.476 \ \ M_\odot \nonumber
\end{eqnarray}
The top panels of Fig.~\ref{fig:pca}
show the best edge-on views of the plane, and the bottom panel
its face-on view.

The first eigenvector PC1 is weighted roughly equally between SFR and
\mstar; its dependence on O/H is relatively small.
The same is true for PC2 with the relative weights between SFR and \mstar\ reversed,
but with a slightly increased dependence on O/H.
This means that most of the dispersion in the 3-space is governed by the spreads
in stellar mass and SFR, rather than by metallicity.

The eigenvector with the least variance, PC3,
is dominated by metallicity with a secondary dependence on \mstar.
Of the three parameters, metallicity is the least influential
on the sample variance. 
Apparently, the plane is defined mainly by SFR and stellar mass \mstar\ and its
vertical extent or thickness by metallicity.
However, this result
may stem from the breadth and coverage of parameter space by our sample.
In fact, our bootstrapping experiments suggest that the PCA coefficients depend 
slightly on the composition of the sample. 
The sample spread in SFR is particularly important as the SFR coefficients can
change by 20-30\%.
Repeating a PCA on different samples would be worthwhile in order to gauge the  
generality of our results and the reliability of the coefficients 
(see below).

Since PC3 contains only $\sim$2\% of the variation among the galaxies, it is
a potentially powerful tool for establishing 
an optimized view of the parameter space defined by metallicity, stellar mass, and SFR.
The tightness of PC3 and its dominant dependence on metallicity
makes it possible to formulate a correlation between O/H,
and some linear combination of the other two variables.
The PCA has essentially distilled the information given by the MZR and the
SFMS (and the trend of SFR with O/H), thus providing the combination of the relations with the lowest dispersion.
If we take PC3 as shown in Fig.~\ref{fig:pca} and set it equal to zero,
we can derive a best-fit relation among O/H, SFR, and \mstar, which is
characterized by small variation over a wide range of parameter space.
This is shown in the left panel in Figure \ref{fig:fp} where we have plotted \logoh\ vs.
the equation that results from equating PC3 to zero
(only galaxies with \mstar $< 3\times10^{10}$\,\msun\ are shown).
\begin{equation}
12+\log(O/H) = -0.12\,{\rm log(SFR)} + 0.33\,{\rm log(M_{\rm star})} + 5.20 \ \ 
\label{eqn:fp}
\end{equation}
\noindent
The equality line in the left panel and the residuals in the right
panel show that the observed abundance is well estimated by the FP.
The reliability of this estimate is roughly constant over
the entire parameter space covered by our sample; 
the errors shown in the right panel of Fig.~\ref{fig:fp} are
well approximated by a Gaussian with a $\sigma\,=\,0.17$,
or $\la$48-50\% uncertainty (0.17\,dex).

Also shown in Fig. \ref{fig:fp} is the FP applied
to the SDSS sample used to define 
the FMR \citep{mannucci10}; only galaxies with \mstar\ $\leq 3\times10^{10}$\,\msun\ are plotted.
The dispersion of the SDSS sample around the FP defined here is small,
$\sigma$\,=\,0.06\,dex, the same as \citet{mannucci10} found relative to the FMR.
However, the SDSS sample appears to be offset from the FP by $\sim$0.4\,dex,
as shown in the right panel of Fig. \ref{fig:fp}.
The mean metallicity of the FMR-SDSS sample (with 55\,082 galaxies having masses
below $10^{10.5}$\,\msun) is 8.94 ($\pm$0.14), roughly the limit of the maximum
oxygen abundance in spiral galaxies \citep[8.95:][]{pilyugin07}.
Such metallicities are also
well into the metallicity range known to have large discrepancies ($\simgt0.4$\,dex)
relative to direct calibrations \citep[e.g.,][]{kewley08,pilyugin10}. 
This is not a problem peculiar to the SDSS sample, since the metal-rich galaxies we examine
here fall in a similar metallicity region (see Fig. \ref{fig:fp}).

As stated by \citet{moustakas10}, ``the factor of 5 absolute uncertainty in the nebular abundance
scale poses one of the most important outstanding problems in observational astrophysics".
There is some evidence that calibrations based on theoretical strong-line calibrations
\citep[e.g.,][]{tremonti04,kobulnicky04,nagao06}
{\it overestimate} the nebular metallicity by 0.3$-$0.5\,dex \citep[e.g.,][]{bresolin09},
but direct methods may {\it underestimate} it by 0.2$-$0.3\,dex \citep[e.g.,][]{przybilla08}. 
Such discrepancies may be due to temperature mixing in 
the \hii\ regions in metal-rich giant spiral galaxies \citep[e.g.,][]{peimbert67}, 
but are much less common in metal-poor dwarfs \citep[e.g.,][]{pilyugin10}.
In any case, different metallicity calibrations make comparing samples 
over wide ranges of abundances quite problematic, and are ultimately a major
obstacle to probing scaling relations over a vast parameter space.

The FP fits the FMR-SDSS sample as well as the FMR itself ($\sigma\,=\,$0.06\,dex),
although with a $+0.4$\,dex offset which could be due to problems with the metallicity calibration
at high abundances \citep[e.g.,][]{bresolin09}.
The FP dispersion of 0.17\,dex for our sample(s) is higher than that found by \citet{tremonti04} for the MZR
defined by 53\,000 galaxies from the SDSS (0.1\,dex), and also higher than the FMR (0.06\,dex)
found for the similar SDSS sample by \citet{mannucci11}. 
However, both SDSS samples 
span limited ranges in O/H, SFR, and stellar mass relative
to the parameter space covered by our sample.
Indeed, the mean and standard deviation
of (log) stellar mass for the FMR-SDSS sample (with \mstar$< 10^{10.5}\,\msun$)
is 10.07$\pm$0.31 (the comparable mean, standard deviation in our sample is 8.48$\pm$1.10).
Thus,
a lower dispersion in the fit is not surprising despite the many more galaxies in those samples.
The value of the FP dispersion for the $\sim$1000 galaxies studied here is only slightly higher than
that found for the MZR of 25 nearby dwarf galaxies \citep[0.12\,dex,][]{lee06}, another
sample dominated by low-mass galaxies.
\citet{tremonti04} and \citet{mannucci10} found that dispersion in the MZR increases with decreasing mass;
hence given the parameter space covered by our sample, the $\sigma$ of 0.17\,dex is
reasonable.

It is also true that our sample is rather heterogeneous, in terms of the methods
with which metallicity and stellar mass are calculated.
Unfortunately this is of necessity since we are comparing properties
of local and high-redshift galaxy populations, over a wide range
of metallicity.
Thus, it is possible that intrinsic systematics could be responsible for a fraction of the 
dispersion observed for our sample.

\subsection{PCA of the SDSS sample}\label{sec:pcasdss}

Because our original aim was to preferentially select metal-poor starbursts, our combined
sample could be biased by metallicity.
In such a case, the FP found here would not be applicable to general galaxy populations.
To test this possibility, we performed a PCA of the 
SDSS sample used by \citet{mannucci10}, considering only those galaxies
(55\,082) with \mstar$\leq 3\times 10^{10}$\,\msun, the same limit as imposed for our sample.
This is the same FMR-SDSS sample discussion in the previous section.
If the coefficents and dispersions are similar to the ones obtained from the PCA on our sample,
we could conclude that our sample is not unduly biased, and that the results
can be generalized to broader galaxy populations.

\begin{figure*}
\hbox{
\includegraphics[width=0.5\linewidth,bb=18 144 592 718]{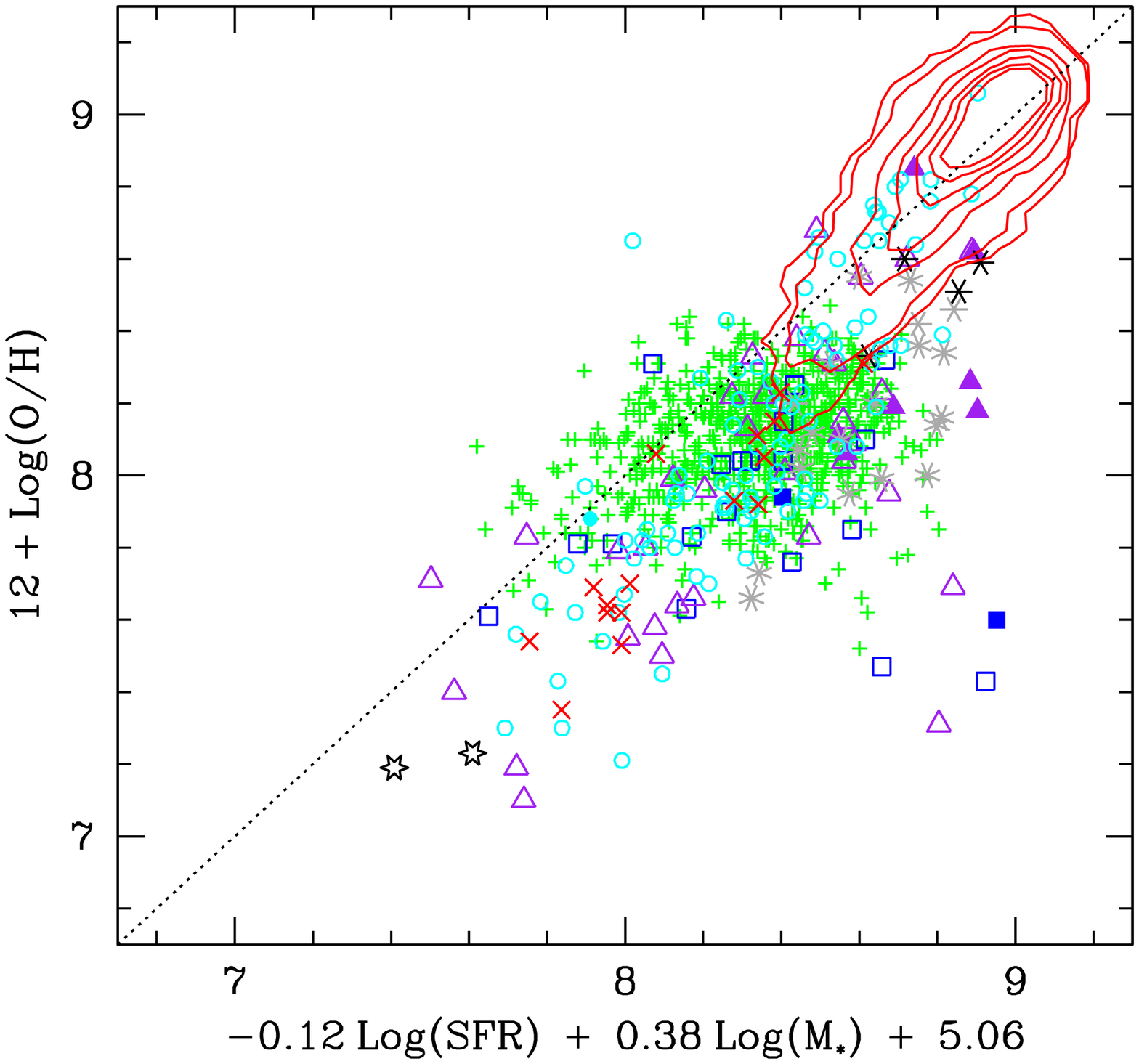} 
\includegraphics[width=0.5\linewidth,bb=18 144 592 718]{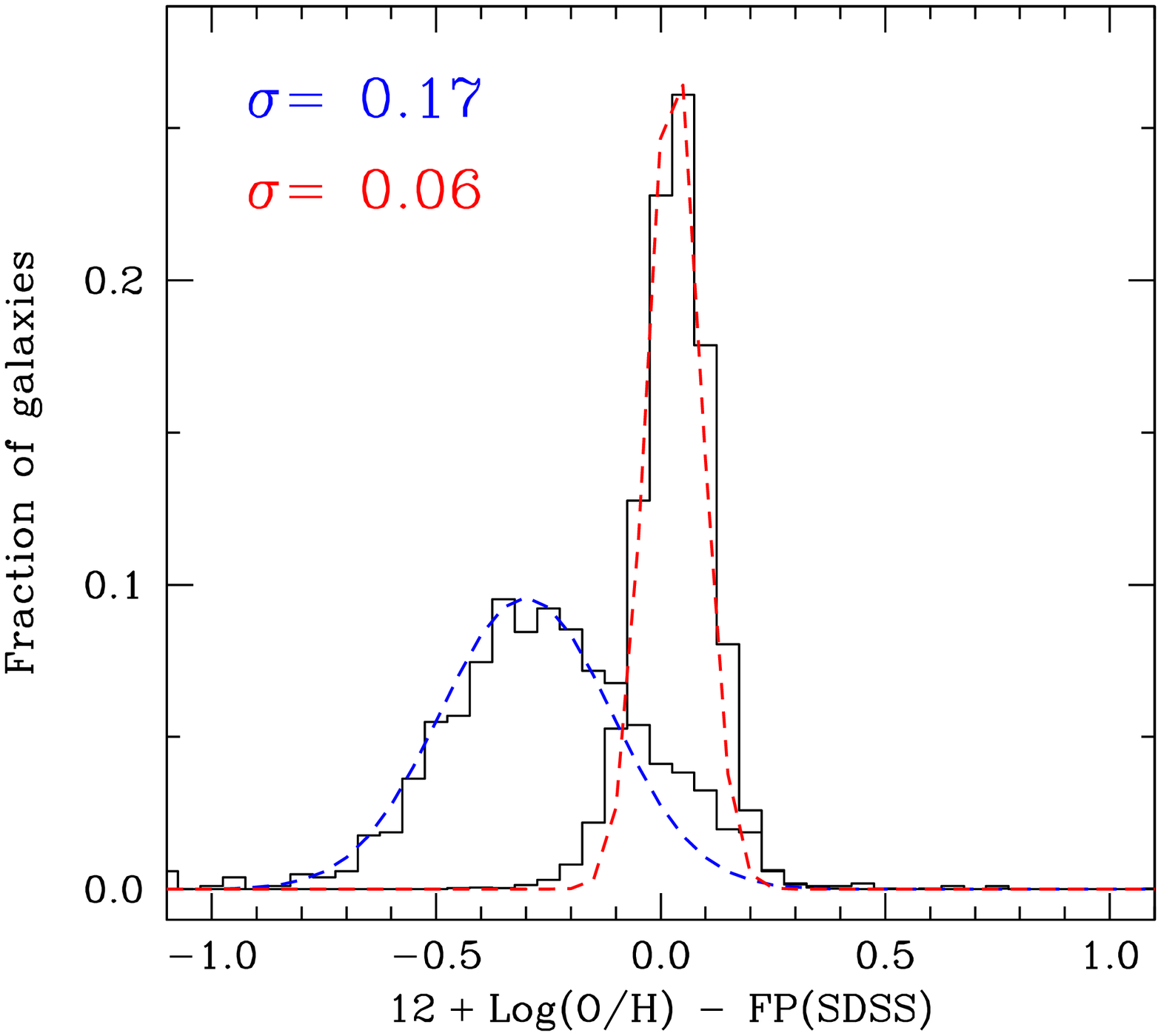} 
}
\caption{Left panel: the PC3 component shown as \logoh\ vs. the resulting 
linear combination of (log) SFR and \mstar\ obtained from the PCA on the FMR-SDSS sample.
Galaxies are coded as in Fig.~\ref{fig:pca}.
As in Fig.~\ref{fig:fp}, the FMR-SDSS sample is shown as red contours,
with the innermost contour corresponding to 60\% of the sample of 55\,082 galaxies,
and increasing to 70\%, 80\%, 90\%, 95\%, 99\%, with 99.5\% as the outermost contour.
The dotted line in the left panel indicates equality.
Right panel: the histogram of the residuals from the identity (dotted line)
relation shown in the left panel;
the blue Gaussian and broader distribution corresponds to our sample,
and the red Gaussian and narrower distribution to the FMR-SDSS galaxies.
The distribution of residuals for our samples is well fit by a Gaussian but offset by $-0.3$\,dex.
In both samples (and both panels), only galaxies with \mstar\ $< 10^{10.5}$\,\msun\ are shown.
}
\label{fig:fpsdss}
\end{figure*}

The eigenvectors resulting from the PCA on the FMR-SDSS are found to be very similar 
to those we obtained for our sample.
Specifically, 98\% of the variation among the parameters
is contained in the first two eigenvectors, PC1 and PC2,
and they are similarly dominated by \mstar\ and SFR.
Moreover, PC3, dominated by O/H, comprises only 2.3\% of the variance. 
Thus, following the same reasoning as before, we have set PC3 equal to zero
and solved for \logoh.
From the PCA on the 55\,082 galaxies in the mass-limited FMR-SDSS, we find:
\begin{equation}
12+\log(O/H) = -0.12\,{\rm log(SFR)} + 0.38\,{\rm log(M_{\rm star})} + 5.06 \ \
\label{eqn:fpsdss}
\end{equation}
\noindent
Comparison with Eq.~\ref{eqn:fp} shows that the
coefficient multiplying (log) SFR is the same as for our sample,
but there is an offset of $-0.14$ in the constant. 
The \mstar\ coefficient is also slightly steeper than for our sample (0.38 vs. 0.33), 
although not as steep as the 0.51 coefficient multiplying \mstar\ in the
extension of the FMR to low mass by \citet{mannucci11}.
The difference in the FP constants ($-0.14$) is not as large as the offset
between the two samples ($-0.4$, see Fig.~\ref{fig:fp}),
but the larger coefficient multiplying (log) \mstar\
compensates the difference.
Fig.~\ref{fig:fpsdss} shows both samples plotted with
the FP defined by the FMR-SDSS galaxies,
and the relative dispersions. 

If we apply this new (SDSS) FP to our sample, we find the same spreads
for both samples as with the original FP
(0.17\,dex, 0.06\,dex, see Fig.~\ref{fig:fpsdss}).
However, our sample requires an offset of $-0.3$\,dex, slightly smaller 
(and of opposite sign as expected) than
the SDSS offset relative to the original FP ($+0.4$\,dex).
Hence, the two formulations are probably equally good representations of
galaxies up to $z\simgt3$,
across a wide range of stellar mass ($10^{6}<$\mstar$<3\times10^{10}$\,\msun),
metallicity, and SFR ($10^{-4}<$SFR$<10^{2}$\,\msunyr).
Nevertheless, as discussed above, differences in metallicity calibrations make
it difficult to accurately assess scaling relations across wide
ranges of abundances.

\subsection{Comparison with the FMR}\label{sec:comparison}

\begin{figure*}
\hbox{
\includegraphics[width=0.5\linewidth,bb=18 144 592 718]{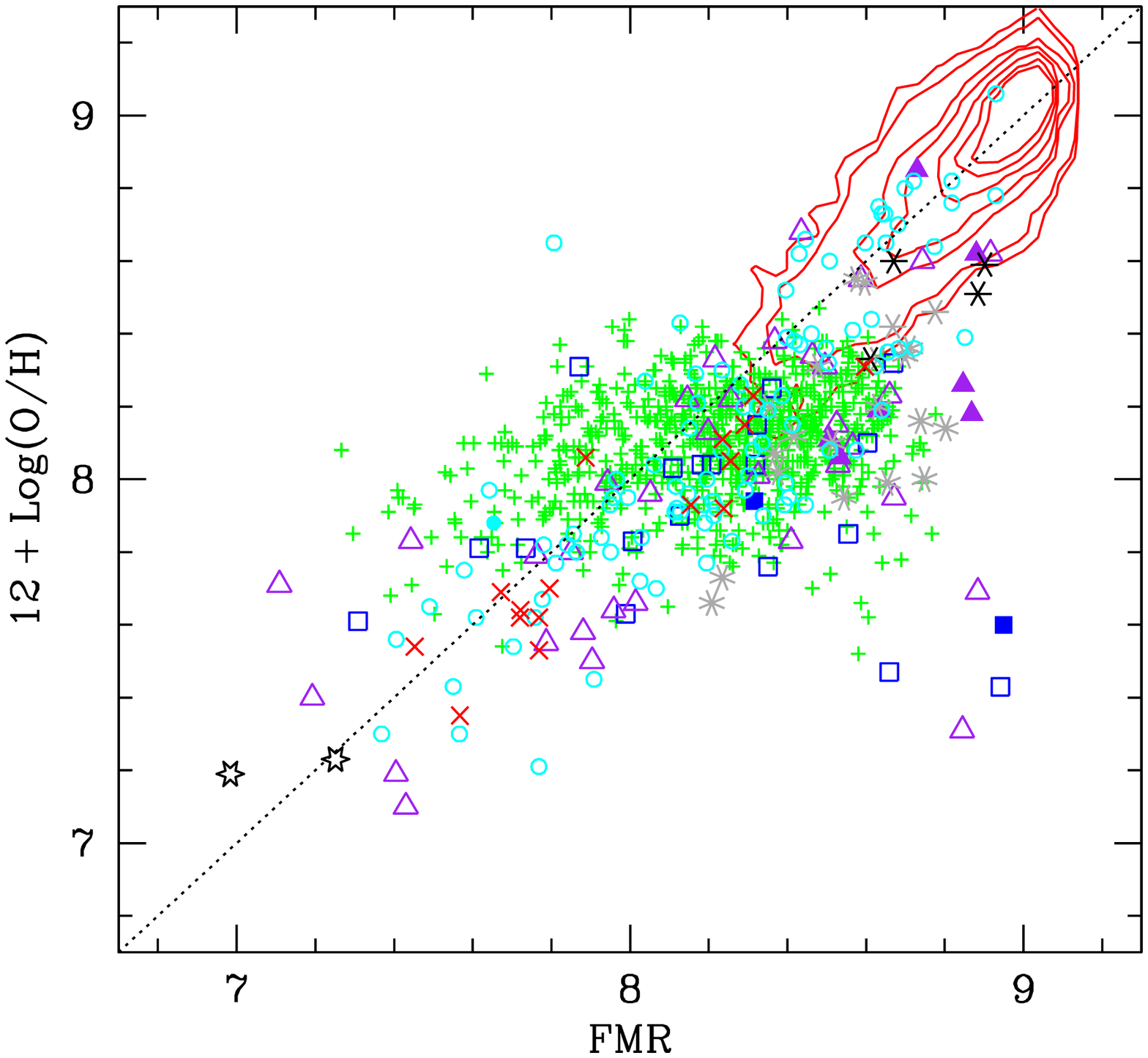} 
\includegraphics[width=0.5\linewidth,bb=18 144 592 718]{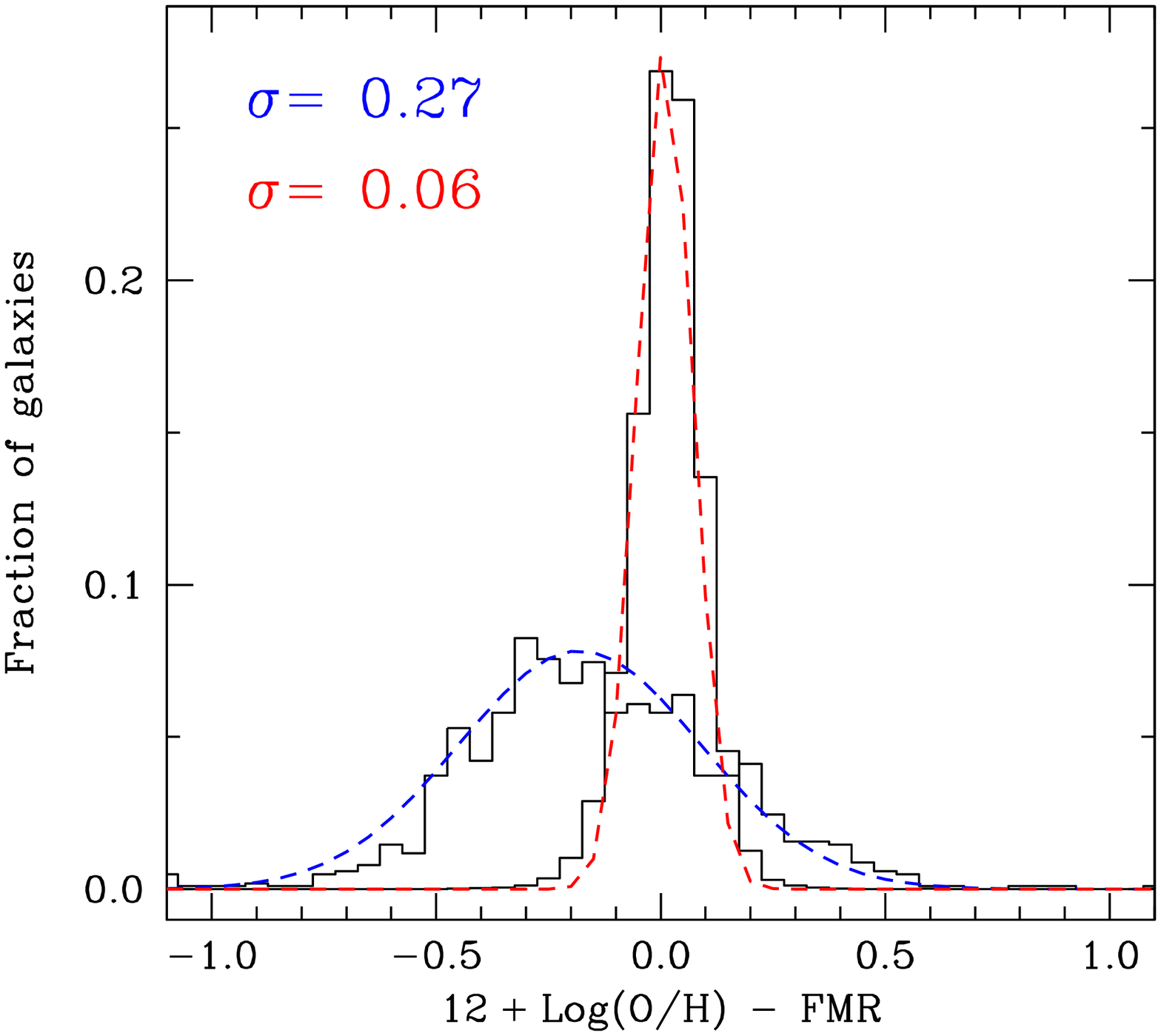} 
}
\caption{Left panel: nebular oxygen abundance, \logoh, plotted against the FMR abscissas,
as described in the text \citep[there are two separate branches, according to the
value of $\mu_{0.32}$, see][]{mannucci11}.
Symbols and contours are as in Fig.~\ref{fig:fp}; the dotted line indicates equality. 
Right panel: the histogram of the residuals from the identity (dotted line)
relation shown in the left panel;
the blue Gaussian and broader distribution corresponds to our sample,
and the red Gaussian and narrower distribution to the FMR-SDSS galaxies.
Our sample is offset by $-0.18$\,dex, with a slightly asymmetric distribution.
In both samples (and both panels), only galaxies with \mstar\ $< 10^{10.5}$\,\msun\ are shown;
this mass limit is clearly evident in the left panel.
}
\label{fig:fmr}
\end{figure*}

As mentioned in the Introduction,
\citet{mannucci10} devised the Fundamental Metallicity Relation
which significantly reduces the scatter in the MZR of SDSS-selected galaxies
by introducing a dependence on the SFR.
\citet{mannucci11} extended the FMR to lower masses 
($\mu_{0.32}<9.5$)
such that:
$$12+\log(O/H) = 0.51(\mu_{0.32}-10)+8.93 $$
\begin{equation}
\quad\quad\quad\quad = -0.16,{\rm log(SFR)} + 0.51\,{\rm log(M_{\rm star})} + 3.83 \\
\label{eqn:fmr}
\end{equation}
\noindent
where $\mu_{0.32}$\,=\,log(M$_{\rm star}$)$-$0.32\,log(SFR).

We have applied the FMR to our sample (considering also the non-linear
extension to high masses\footnote{\logoh\,=\,$8.90+0.37m - 0.14s - 0.19m^2
+0.12ms - 0.54s^2$ for $\mu_{0.32}\geq9.5$,
where $m=\log(M_{\rm star})-10$ and $s=\log$(SFR) in solar units.
See Eq. 2 in \citet{mannucci11}.}) 
and find that the residuals have mean values of 
$\ga$0.32\,dex,
almost twice the value we find here for either FP. 
The FMR applied to our sample and to the FMR-SDSS sample is shown in Fig.~\ref{fig:fmr}.
The residuals for our sample are asymmetrically skewed to negative values, and
fitting them to a Gaussian gives $\sigma\,=\,0.27$\,dex. 
Hence, the FMR is a poorer fit to our sample than the FP.

The reason is that the extension to lower masses proposed by \citet{mannucci11}, with a slope
of $0.51$, is too steep to fit the galaxies in our sample
(see also Fig.~\ref{fig:sfr_oh}).
From the PCA, we find a metallicity-mass dependence of 0.33, similar to
\citet{lee06} who found for their low-mass dIrr sample a linear dependence (in log-log)
of 0.30 between \logoh\ and \mstar\ ($\sigma\,=\,$0.12\,dex).
This shallower dependence of metallicity on mass may be peculiar to samples 
with many low-mass 
galaxies, although we find similar values even when restricting the fits to 
only the highest masses,
and when fitting the FMR-SDSS sample.

Nevertheless, the relative dependence between \mstar\ and SFR found by our PCA is similar to
that given by \citet{mannucci10}.
With $\mu_{0.32}$,
they assumed a unit dependence of O/H on (log)\mstar, and
found that the value of 0.32 multiplying (log)SFR minimizes the 
residuals of the MZR relative to $\mu$.
The extension to lower masses by \citet{mannucci11} 
has a flatter dependence of O/H on $\mu_{0.32}$, 0.51 (see above),
but the same value of 0.32 relating (log)SFR to \mstar.
For the FP,
we find a shallower dependence of O/H on \mstar\ (0.33), and a SFR coefficient of
$-0.12$; multiplying 0.33 (our mass dependence) by the 0.32 SFR factor of
\citet{mannucci10} would give a SFR dependence of $-0.11$, very similar
to the value of $-0.12$ given by the PCA. 

\section{Discussion and conclusions}\label{sec:conclusions}

From a sample of $\sim1100$ galaxies spanning a wide range of O/H, SFR, and \mstar,
we have identified a class of low-metallicity starbursts which deviate from the SFMS
and the MZR in the same way, independently of redshift.
When these scaling relations are considered together, 
the sample as a whole (although excluding the galaxies with 
\mstar$>3\times10^{10}$\,\msun) defines a Fundamental Plane whose orientation is defined
primarily by SFR and \mstar, and whose thickness is governed by O/H.

The two-dimensional nature of the plane in the 3-space defined by
the observables implies that only two parameters are necessary to describe
a galaxy \citep[see also][]{laralopez10}.
The thinness of the plane when viewed edge-on enables the formulation
of a relation which combines the SFMS and the MZR in an optimal way.
Through application of a PCA, we have established a dependence of O/H
on SFR and \mstar\ which has a deviation over the whole sample 
(for \mstar$\leq3\times10^{10}$\,\msun) of 0.17\,dex, or $\sim$48\%.

The FP from the PCA applied to
the SDSS sample used to define the FMR \citep{mannucci10} is similar
to that obtained for our sample
(see Eqs.~\ref{eqn:fp}, \ref{eqn:fpsdss}). 
Applying either FP to our sample and to the FMR-SDSS
shows an offset in metallicity which distinguishes the two, but
the dispersions are similarly low
(0.17\,dex and 0.06\,dex for our and the FMR-SDSS samples, respectively). 
However, the FMR is not the best description of our sample, as it
gives a higher dispersion (0.27\,dex) than either FP.

Perhaps the most intriguing aspect of the FPs given by the PCA
is that they hold for all the redshifts we have examined, $z\la3$.
In our sample, the LBGs at all redshifts follow the ``Fundamental Plane''
relation which combines the MZR and the SFMS. 
{\em There are no high-redshift outliers.}
Moreover, the low-metallicity starbursts at all redshifts also follow
the relation.
On the basis of O/H, SFR, and stellar mass alone, apparently there is
no distinction between the behavior of galaxies in the Local Universe
and those at high redshift (at least up to $\la3$).

The statistical analysis does not clarify which relation or which 
observable is the most important; it merely indicates that there is 
a {\em plane} that describes the family of
galaxies in the 3-space of O/H, SFR, and \mstar.
However, the small amount of variance contained in the eigenvector
dominated by metallicity (PC3) suggests that metallicity is the least
important of the three parameters in driving the variance of our sample.
The physics of why or how these parameters conspire to form a plane
independent of redshift
will be explored in a companion paper \citep{magrini12}.
There we show that the deviations from the main
scaling relations are due to a different mode of star formation,
a starburst or ``active'' mode which evolves over relatively short timescales.
The main scaling relations are instead defined by a more quiescent,
``passive'' mode of star formation which evolves more slowly.

The existence of a Fundamental Plane that spans redshifts up to $z\ga3$
implies that the SFMS and MZR do not really change with lookback time, 
but rather that the galaxy populations that define them change with redshift.
Starbursts become increasingly common at high redshift,
but they also exist locally and at low metallicity. 
Thus, the ``evolution'' in the scaling relations
is a result of selecting those galaxies which are most common
at a particular redshift,
and may not reflect a true change in the physics of how galaxies evolve.

\section*{Acknowledgments}
We warmly thank the 
International Space Science Institute (ISSI) for financial support for
our collaboration,
MODULO (MOlecules and DUst and LOw metallicity), so that we could meet and
discuss the work for this paper.
We are also grateful to Yuri Izotov, 
Giovanni Cresci and Filippo Mannucci
for passing us their data in electronic form.
L.M. acknowledges the support of the ASI-INAF grant I/009/10/0,
and L.H. and R.S. a grant from PRIN-INAF (2009).

\label{lastpage}

\end{document}